# Digital Building Twins and Blockchain for Performance-Based (Smart) Contracts

Jens J. Hunhevicz[1†], Mahshid Motie[1,2], Daniel M. Hall[1]

[1] ETH Zurich, Institute of Construction and Infrastructure Management, Chair of Innovative and Industrial Construction, Zurich, Switzerland

[2] Basler & Hofmann AG, Zurich, Switzerland

[†] Corresponding author: hunhevicz@ibi.baug.ethz.ch

## Abstract

Servitization business models can use performance-based contracts to consider overall life-cycle costs rather than just production costs. However, practical implementation of performance contracts has been limited due to challenges with performance evaluation, accountability, and financial concepts. As a solution, this paper proposes the connection of the digital building twin with blockchain-based smart contracts to execute performance-based digital payments. First, we conceptualize a technical architecture to connect blockchain to digital building twins. The digital building twin stores and evaluates performance data in real-time. The blockchain ensures transparency and trusted execution of automatic performance evaluation and rewards through smart contracts. Next, we demonstrate the feasibility of both the concept and technical architecture by integrating the Ethereum blockchain with digital building models and sensors via the Siemens building twin platform. The resulting prototype is the first full-stack implementation of a performance-based smart contract in the built environment.

*Keywords: Digital Twin, BIM, IoT, Sensors, Blockchain, Smart Contracts, Performance based contracts, Servitization, Built Environment as a Service*

## 1. Introduction

The global building and construction sector is a major contributor to global energy consumption [1]. Despite governmental efforts to lower energy use and emissions, the trend is still rising [2]. One untapped possiblity for emission reduction is the construction of more sustainable buildings with better lifecycle performance [2]. However, these buildings suffer from the so-called building-energy performance gap, where the actual building life-cycle energy performance does not match predictions [3,4]. Despite the push for more innovative and energy-efficient designs [5], the actual energy usage can be up to 250% higher than the predicted energy usage [6].

Although some root causes for the building-energy performance gap can occur at the design stage – (e.g., miscommunication among stakeholders, poor technology performance, or incorrect simulation models [4]), the construction and operations stages are also at fault. Energy performance can suffer from poor quality of initial construction and or poor operation of the building [4] resulting from organizational and behavioral factors [3]. The final construction quality of the building might not be in accordance with the specification (e.g., poor attention to insulation and airtightness) [7]. Ad-hoc construction solutions can deviate from specified designs and result in unintended consequences that lower performance [7]. Further problems occur during the actual operation of the building. For example, occupant behavior and thermal comfort levels can deviate from assumptions and control settings can be manually changed by the facility management (FM) [4]. Overall operational performance can suffer from a lack of continuity in monitoring, analysis, and control throughout the building lifecycle [4].

Such explanations for the building-energy performance energy gap illustrate the role of localized decisions and self-interested actions commonly found in the highly-fragmented architecture, engineering, and construction (AEC) sector. AEC suffers from a misalignment of incentives across the different stakeholders and life-cycle phases, which hinders holistic and systemic innovations [8]. The different set of stakeholders, decision-makers, and values in each phase creates displaced agency – also called "broken agency" – where involved parties will engage in self-interested behavior and pass costs and risk off to others in the supply chain in subsequent life cycle phases [9]. Furthermore, the prevalent low-bid culture in construction also favors low-cost solutions at the tendering stage over solutions that minimize costs over the whole building life cycle [10].

To address this, performance-based building has been recognized as a promising solution [11]. Performance-based building contracts are legal instruments intended to financially incentivize parties to deliver a building that meets targeted performance levels. These contracts bind the profit of parties to longer-term commitments based on mutually determined baseline performance levels during operations [12,13]. Performance-based contracts in the built environment can be understood as a new and compelling business case [14] called *servitization*. Servitization – also known as "Product-as-a-Service" – is a business model embraced by the manufacturing industry where products are leased out to the customer on performance contracts, while still being operated, maintained, and recycled by the producer [15]. Servitization offers competitive advantages to the producers, lets customers profit from higher quality products and services, and benefits the environment through more reuse, recycling, and dematerialization [15–18]. Implemented in the built environment through performance-based contracts, servitization can align incentives over the life cycle of a building and address the energy performance gap [3].

However, servitization using performance-based contracts has not been widely adopted in the built environment [14]. Scholars note issues with accountability [11], the lack of standardized performance evaluation [19], new and unfamiliar financial concepts [19], and the burdens of additional upfront communication efforts between parties [12]. Trial projects (e.g., the private finance initiative [20] in the United Kingdom) promote the idea of the "built environment as a service" but have not generated much traction. The standard practice remains that building owners pay designers and builders a capital sum for initial construction while bearing themselves the long-term risk that comes from operating and maintaining the assets, even when they do not meet promised performance requirements.

However, the ongoing digitalization of the industry and new technologies like digital twins and blockchain present a new opportunity to better implement performance-based building [21]. The rise of digital building twins creates a bi-directional link between physical reality and the digital replica of a built asset [22]. The digital twin concept is widely used in manufacturing to accurately reflect the real-world state in a virtual model. At the same time, the digital twin can adjust the real-time behavior of the physical product according to the performance assessments of the virtual model [23]. Digital twins can enable performance-based contracting by setting performance expectations through simulation, measuring and updating the actual state of performance, and providing recommendations for operations and maintenance through analytics. Overall, digital twins can help to predict and measure performance accurately and equitably, thus overcoming a noted barrier to performance-based energy contracting in the built environment [13].

Furthermore, blockchain can ensure an unchangeable and transparent digital record of transactions. Some blockchains also support the execution of scripts called smart contracts to define tamperproof transaction logic. A fundamental challenge for performance based building is accountability [11], an issue that blockchain can address by ensuring protection mechanisms that help to avoid the risks and costs of opportunistic behavior in construction supply chain collaboration [24]. However, to date, few attempts have been made to study or to implement performance-based smart contracts.

This paper investigates how blockchain based (smart) contracts in combination with digital building twins could support a transition to a more performance-driven built environment.

## 2. Departure

### 2.1. Towards a performance-based built environment

Product-as-a-service business models have been successful in the manufacturing industry [15]. Adapting this model, Figure 1 conceptualizes the difference between a traditional and a servitized business model in the built environment.

In traditional construction, the owner usually pays a capital sum for the delivery of a built asset such as a building. This price includes the construction and initial commissioning of the project. Over the lifecycle of the asset, the owner is responsible for financing the operation, maintenance, and disposal of the asset (Figure 1, a)). This gives little incentive to contractors to design and build for the best possible life cycle performance, as they are not involved in later phases and their reward does not depend on life cycle performance.

In a performance-based building, the user would only pay for the provided services. Ownership and responsibility for operations, maintenance, and disposal stay with the producer (Figure 1, b)). This aligns the interest in designing and building for the best possible performance with the interest in minimizing operational, maintenance, and disposal costs (e.g. through recycling and reuse) in order to maximize profits.

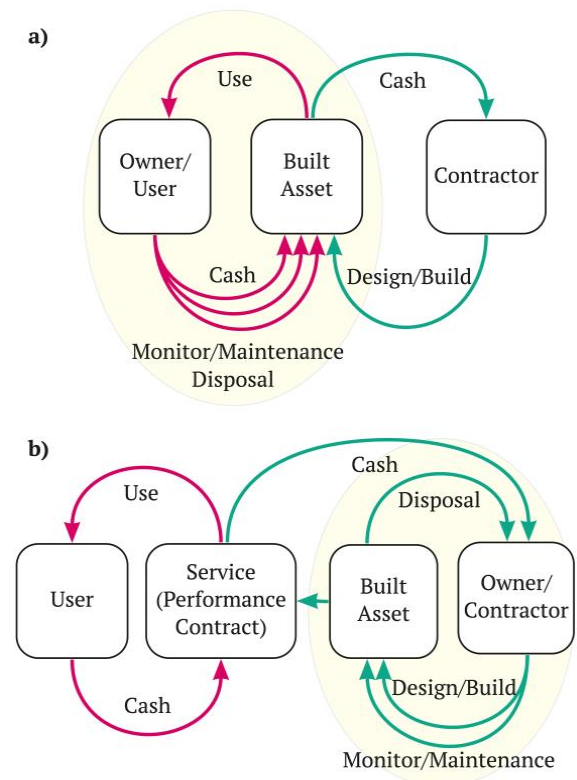

*Figure 1: (a) Traditional payment of a capital sum for the building at the end of construction. The user is responsible for operations, maintenance, and disposal. (b) In a built environment as a service, the user purchases the agreed services provided by the built asset. The owner/contractor takes care of production, operations, maintenance, and disposal. (Adapted from [15])*

Figure 1 is a simplification and does neglect many differences between the built environment and manufacturing. A built asset consists of numerous sub-products that provide different services. Also, more stakeholders might be involved, e.g. the owner could still be an investor rather than the producer. Nevertheless, the core message remains unchanged: servitization aligns interests across the asset life-cycle to maximize performance [15–18]. This is true regardless of which servitized asset or which stakeholders participate in the performance contract. Furthermore, it would be possible to only servitize certain technical sub-systems instead of whole buildings (e.g. heating or lighting) [21].

This paper focuses on the potential role of digital building twins and blockchain-based smart contracts to enable digital, trusted, and automated performance contracts as the central element towards a servitized built environment (Figure 1, b)).

### 2.2. Current practice of performance-based contracts

Performance-based contracts link the building contractor or supplier to a longer-term commitment beyond the initial construction and handover of a facility [12]. To form the contract, parties mutually agree on a baseline performance level as the reference for determining the returning profits [13]. Performance contracts can unite a building owner with a building contractor/operator for a shared profit goal [25]. As a specific example, an energy performance contract establishes a link between building equipment and energy performance gains. The payments provide builders and operators with a long-term incentive to maintain and improve equipment performance [26,27]. This contrasts with contractors in a conventional project, who are not involved in operations or maintenance and have no incentive to improve equipment performance after installation [26].

It is increasingly necessary to link construction project management to building performance and in particular to environmental sustainability performance [28]. For example, Papachristos et al. [28] use a system dynamics model combining project management and building energy performance to demonstrate that intra- and inter-stage partner alignment can increase building performance quality by 6.3%.

However, overall progress in the adoption of such performance contracts remains slow [19]. Pätäri and Sinkkonen [19] identify several risks and barriers to implementing performance-based contracts that are relevant to our study. Financial challenges include a lack of appropriate forms of finance due to conservative lending practices, limited experience in understanding performance-based project financing, a lack of confidence in servitization contracting, and a lack of standardized measurement and verification procedures for performance savings. Additional challenges come from the increased duration and complexity of the communication between the contractor, the client, and the tenants and building users [12], as well as from issues of accountability in the case of performance failures [11]. Performance-based contracts might require contractors to re-examine business models, exploring vertical integration or direct employment to provide continuity of care over their completed buildings [12,19]. Scholars have called for exploration of how the new business models and new financing models of performance contracts can be combined with emerging automation technologies such as digital twins and the internet of things (IoT) [29], but little research to date has explored this in detail.

### 2.3. Digital Building Twins

The digital twin is a virtual replica of a physical asset [30]. The concept of digital twins requires three parts: the physical product, the virtual replica, and the linkage between them [31]. The linkage is achieved using the IoT, which describes the concept of devices (things) with embedded electronics and software that collect and exchange data through the internet [32]. In the digital twin concept, such smart devices collect data and transmit it to the virtual representation in the cloud, but also vice versa to optimize the physical product state based on analytics conducted on the virtual model [23]. Digital twins are understood as one of the key enablers of digital transformation in the manufacturing industry [23,31,33,34]. While already adopted in many cases, research in manufacturing still investigates how the real-time integration of IoT and simulations can be improved [35–37].

As in manufacturing, digital building twins are envisioned as the next big step towards a digital construction and built environment, allowing for real-time performance optimization of built assets [22,38,39]. The adoption of building information modeling (BIM), which is the continuous use of digital building models throughout the lifecycle of the built facility [40], is seen as the basis for this transformation. In contrast to digital twins, most digital building models still do not include any form of automated data exchange between the physical object and the digital object. Connecting BIM with IoT allows the digital model to be updated according to changes in the physical state of the building [41–43]. Studies have only recently begun to research the potential of digital twins for performance optimization through real-time assessment of what-if scenarios in the virtual space in construction processes [44,45], sustainability-based life cycle management of railway [46], operations management of HVAC systems [47], and maintenance of bridges [48].

Despite the early research state, digital building twins are commonly seen as the inevitable evolution of BIM concepts towards more integrated and automated life cycle approaches [45] that focus on closing the information loop between digital and physical built assets [39]. They provide a platform to build data-driven and real-time performance-based contracts.

### 2.4. Blockchain

Blockchain is the most common type of Distributed Ledger Technology (DLT) [49–51]. It consists of a distributed record of transactions (called a ledger) in a peer-to-peer (P2P) network, where encoded governance rules incentivize participants to cooperate in adding transactions and securing the network. As a result, a blockchain can ensure an unchangeable and transparent digital record of transactions, which allows anonymous transacting parties to trust each other without intermediaries. For now, cryptocurrencies (e.g. Bitcoin [52]) are the most prominent use case of blockchain. Newer networks innovate on the application layer built on top to enable new use cases through so-called smart contracts. Smart contracts are scripts that encode interaction logic with transactions and run unchangeably on the blockchain. Ethereum [53] was the first blockchain to enable such Turing-complete smart contracts. One of the most prominent smart contract use cases to date is decentralized finance (DeFi), which replicates financial services on the blockchain without the need of financial institutions [54].

Recently published literature reviews reveal a strong increase in publications that investigate blockchain across many sectors and in combination with other technologies [55–57]. Likewise, recent reports [58–60] and articles [61–66] discuss blockchain use cases also for construction and the built environment. Hunhevicz et al. [63] cluster use cases into seven categories and assess with their framework whether a DLT (blockchain) is needed. In brief, blockchain is needed when no third party can or should be involved, as well as when not all participants are known or interests are not aligned. Many of the proposed use cases apply blockchain to existing processes where stakeholders are known, so blockchain might not be necessarily required or at least needs further investigation [63]. However, use cases that involve coins and tokens for new payment or incentive schemes were found to be highly likely to rely on the use of blockchain [63]. Several publications support this observation by investigating blockchain-based payments along the construction supply chain [67–75]. In sum, the use of blockchain in construction promises to increase trust in existing processes through transparent and immutable transactions [24].

The building of new incentive systems with trusted processes and unknown participants has led to new research streams referred to as token engineering or cryptoeconomic design [76]. In the construction industry, the concept of cryptoeconomic incentives has been proposed as means to add a layer of monetary/non-monetary incentives to processes to increase trust and collaboration across life cycle phases and stakeholders [77], e.g. to incentivize high-quality data sets [78]. Performance-based smart contracts seem well aligned with this concept and are therefore likely to benefit from a blockchain.

### 2.5. State of the art

This section reviews the state-of-the-art research in construction and the built environment at the intersection of BIM, IoT, blockchain, and performance-based contracts.

Huang et al. [79] found blockchain in combination with digital twins promising as a means to improve data management. Timestamping transactions helps to keep track of changes, as well as to manage data access, data sharing, and data authenticity among a network of actors. Lee et al. [63] propose that the above can also be promising in construction for accountable information sharing. Their prototype records and timestamps data from a robot sent to its digital twin in near real-time on the blockchain, therefore implementing a full-stack prototype that connects a digital twin with blockchain. They highlight the future potential of automatic payments, but do not discuss or implement any link to performance-based smart contracts. Similarly, Hamledari and Fischer [80] present a full-stack prototype that transmits data from reality capture technologies on-site to a blockchain smart contract, in order to automate payments and the transfer of lien-rights through tokens. Also, Chong and Diamantopoulos [73] present a full-stack protoype that sends data from smart sensors to the BIM model and smart contracts in order to execute secure payment in a façade panel supply chain. Neither of these last studies assesses performance based contracts.

O'Reilly and Mathews [81] propose blockchain and a digital twin to enable financial incentives to design for better building performance during operations. They modeled an imaginary room with four heat sensors and connected it to a dynamo code that fetches the virtual sensor data and stores it in a simulated blockchain environment. However, their prototype simulates both the blockchain and IoT part and does not yet implement the described incentives through a smart contract. Li et al. [82] demonstrate how sensors can gather data on the performance of a simulated installation task, store this data in the blockchain, and use a smart contract to issue automatic payments if the predefined performance conditions are met. However, this prototype does not fully connect IoT, a digital model, and an operational blockchain. While some literature discusses the potential of legal contracts on the blockchain on a conceptual level [83,84], Gürcan et al. [85] developed the first prototype of an energy performance smart contract using the Ethereum blockchain. They successfully tested their smart contract on a private network instance with five-day weather and building performance data set. However, there was no actual connection to sensors and a digital twin, nor was the purpose of the performance contract to incentivize performance across life cycle phases.

### 2.6. Research gap and scope of the study

Although performance-based building has the potential to address the observed energy performance gap, performance-based contracts have not yet been widely implemented. Digital building twins analyze real-time performance data of buildings and can provide a data baseline for performance-based contracts. Nevertheless, the fragmented construction industry faces trust problems across life-cycle phases and trades, and digital building twins alone are unlikely to address this substantially. Blockchain, however, could facilitate trusted cross-phase processes and contracts, building upon the performance data provided by digital twins.

Despite the potential, no research has yet investigated cross-phase performance contracts leveraging blockchain smart contracts and digital building twins to incentivize stakeholders along the built asset life-cycle. Therefore, we illustrate how blockchain smart contracts and digital building twins can interact to enable digital, performance-based contracts. To move beyond theory, we prototype a full-stack architecture using the Ethereum blockchain and the Siemens building twin platform to implement an exemplary cross-phase thermal performance smart contract. The smart contract was successfully tested over two days with sensor data obtained from the digital building twin of a real-world building.

Based on the findings, the paper discusses the challenges and opportunities of performance-based smart contracts in combination with digital building twins to move towards the potential new paradigm of a built environment as a service.

### 3. Proposed performance-based smart contract architecture

We introduce the necessary components to facilitate performance-based smart contracts for a built asset as visualized in Figure 2. A cyber-physical system is characterized by two layers: the physical world and the cyber world. In the physical world, the actual built asset is equipped with sensors that can measure various performance metrics. Furthermore, human stakeholders interact with the built asset, the digital twin of the built asset, or the performance-based contract.

The following section describes in more detail the core components of the cyber world: the building twin platform, the performance-based smart contracts, and the data bridges required for the blockchain we call "front-end oracle" and "back-end oracle".

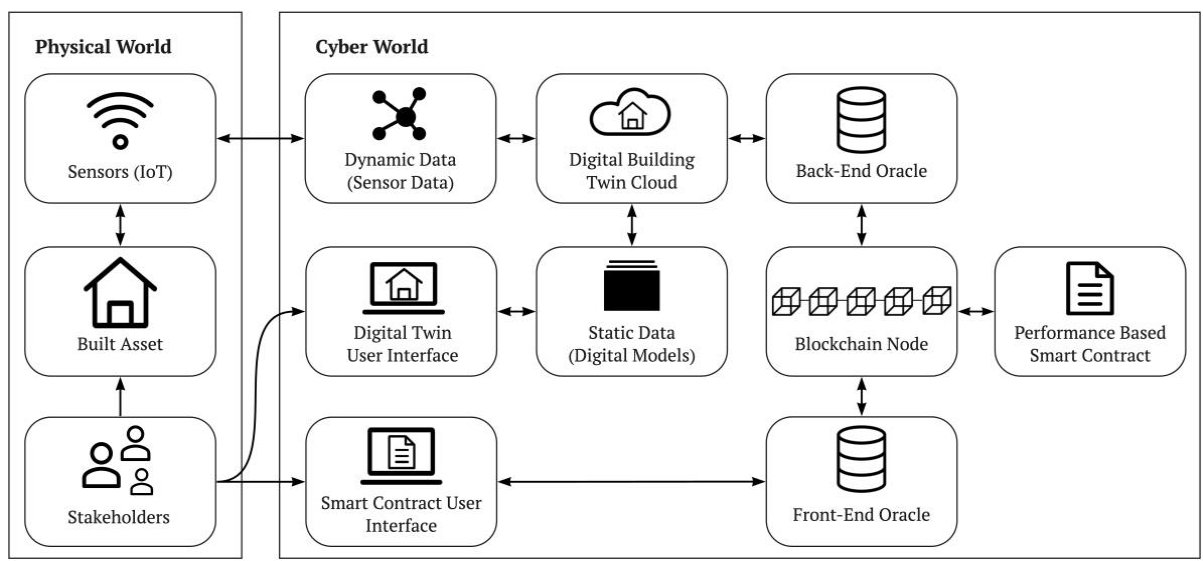

*Figure 2: Interaction of needed cyber-physical components for performance-based smart contracts of built assets.*

### 3.1. Building twin platform

The digital building twin is hosted on one or multiple cloud servers, where data is processed and stored and performance simulations are facilitated. We refer to two types of data: dynamic and static data. Dynamic data refers here to the constant live-data stream captured by the sensors. Static data refers to all other data created by human stakeholders, most importantly the digital BIM models (i.e. IFC files). The stakeholders interact through a graphical user interface with both the static and dynamic data. In most digital building twins, the dynamic data is mapped to a spatial location in the digital model, accessible in the digital twin user interface.

### 3.2. Performance-based smart contracts

The smart contracts created on the blockchain encode the rules of the performance-based contract. Their core functionality can be describe as continuously receiving performance data, checking the data against the encoded contract logic, and executing the subsequent workflow steps (e.g. payments). Figure 3 displays the interaction of needed components that together form a performance based smart contract.

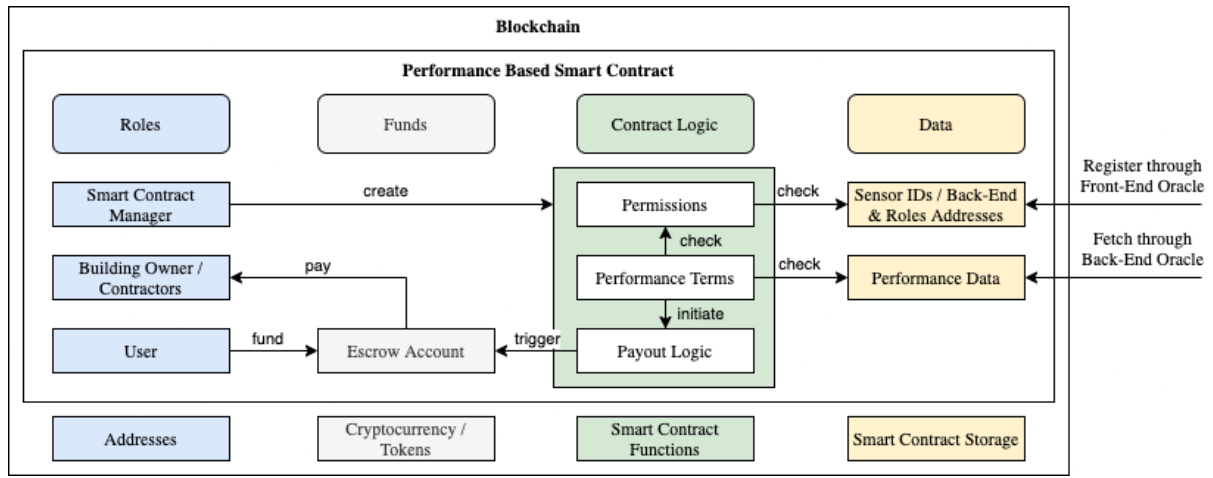

*Figure 3: The interaction of roles, funds, contract logic, and data in a performance based smart contract.*

First, a performance based smart contract must manage the different contracting parties. A smart contract manager is needed settting up the smart contract. Other roles could be the building owner and/or contractors who need to deliver the service, and one or multiple users who receive and pay for the service. So-called roles can then be assigned to blockchain addresses that are allowed to modify and execute the respective transaction. For example, the owner of a smart contract can register for an address "contract_owner", or the owner of the built asset can register for an address "asset_owner". These two roles can have different rights assigned to them for interaction with smart contract functions. It is important to note that one address can also be assigned to multiple roles.

Second, the smart contract encodes the contract logic through smart contract functions. Permissions are assigned to the defined roles and addresses registered in the smart contract. Before executing workflow logic, the smart contract must check whether the address initiating a transaction is allowed to do so. The performance terms then encode the agreed levels of service that are continuously checked against the received actual performance data. If the performance levels are met, the payout logic manages the according payments to the service providers.

Third, funds managed by the smart contract ensure that payments can be timely executed through the use of cryptocurrency or tokens. For that the service users need to pay an upfront payment to the smart contract escrow account.

Lastly, relevant data needed to execute the contract logic is stored within the smart contract. This includes addresses of the users and back-end allowed to interact with the smart contract functions, as well as IDs of the sensors and digital twin. Furthermore, external performance data about the observed real-world events is stored within the smart contract. Since blockchain cannot directly obtain external data, a middleware called an "oracle" is required to create a secure connection between the smart contract and an off-chain data resource. The use of such oracles also introduces the "oracle problem" [86]. In essence, blockchain can verify data integrity on its own ledger and network but it cannot know whether data input by humans or sensors are correct in the first place. This leaves open the possibility that malicious actors try to cheat the system by inputting incorrect data. Every implementation of smart contracts relying on oracles should strive to minimize this possibility

In the case of a performance-based contract for a built asset, two oracles are needed: the "front-end oracle" (see 3.3) as a middleware to connect the web front-end with the blockchain to allow direct stakeholder input, and the "back-end oracle" (see 3.4) to connect the digital building twin platform with the blockchain to fetch performance data.

### 3.3. Front-end oracle

A performance-based smart contract benefits from a connection to a graphical user interface so stakeholders can interact directly with the contract in a convenient way. Stakeholder interaction is required to set up the contract and define the contract logic, register the addresses and roles of the users, register the addresses and IDs of the sensors and digital building twin, interact with the smart contract functions, and check the status of the smart contract (so-called states).

Therefore, the front-end provides a web user interface for the input of static information as well as an oracle middleware to transfer this data to the blockchain and smart contract. These tasks could also be achieved without a graphical user interface, but this complicates the setup, deployment, and interaction with the smart contract considerably. Without a graphical interface, all contract addresses, functions, and parameters would need to be known by all stakeholders interacting with the contract.

### 3.4. Back-end oracle

Once the smart contract is set up, it needs to fetch the performance data of the built asset to assess performance logic. Data from the sensors need to be automatically transmitted to the smart contract. Therefore, a back-end oracle ensures the connection between the digital building twin and the blockchain to transmit the sensor data that has already been processed and stored in the digital twin database. The back-end oracle calls the Application Programming Interfaces (APIs) of the digital building twin database, fetches relevant performance data, translates the data received into the right format, and initiates the transaction.

## 4. Proof of Concept

### 4.1. Use Case

To demonstrate the proposed concept and validate the technical architecture, an exemplary performance-based smart contract was developed and tested on a real-world building in combination with its digital twin.

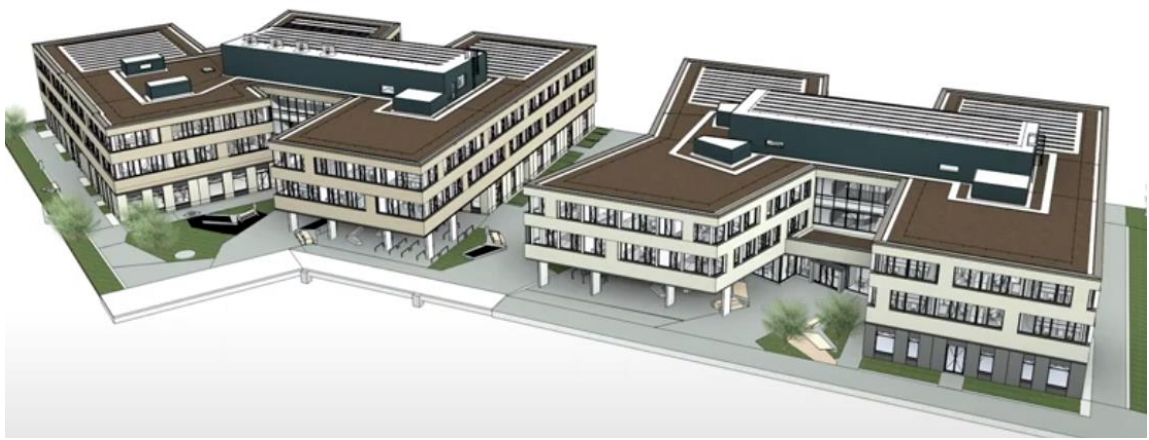

*Figure 4: The IFC model of Tz2 used as a static data source for the digital building twin (Copyright: ATP architekten ingenieure).*

The prototype was tested on the real-world building "Technology Center 2" (Tz2) located in Seestadt, Vienna (Figure 4). It is part of the Aspern Smart City Research center. The commercial building has a floor area of 5600 m2 and can be rented by innovative companies and start-ups. The building is equipped with photovoltaic panels, a heat pump, various energy storage facilities, thermally activated building systems (TABS), smart meters, and sensors. The building condition is monitored and controlled via its digital twin using the Siemens building twin platform.

To limit the scope, we focued on the specific use case scenario of a cross-phase thermal performance contract (see 4.1.2). The full technology stack was implemented, including a front-end oracle with UI to allow participants to set up and input contract information, and a back-end oracle to connect the smart contract to the digital building twin and sensor data. The high-level workflow is depicted in Figure 5.

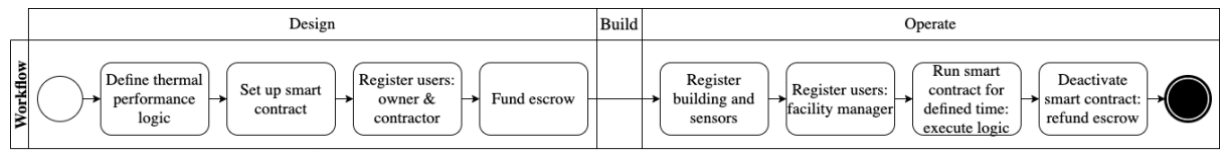

*Figure 5: Work flow for the tested cross-phase thermal performance scenario.*

### 4.1.1.Incentive Design

Since it is a cross-phase performance-based contract, the workflow starts in the design phase (see Figure 5). The scope, logic, and performance basis of the thermal performance contract is defined before the building is constructed. In the implemented use case scenario, the performance contract is established between the building owner, the contractor who designs and constructs the building, and the facility manager that will operate the building. The owner sets the performance target in agreement with the other stakeholders. If the contractor and facility manager meet the performance target, they are paid for the service provided. The smart contract is coded and deployed, and the initial users (owner and contractor) are registered. Instead of paying the contractor a capital sum, the owner funds the smart contract with an escrow to assure that the payments are allocated in the operational phase. Once the building is built and the digital building twin is set up, the building with its sensors and the facility manager role is registered. The smart contract then operates and executes the performance logic and subsequent payments as defined. When not needed anymore, the smart contract is deactivated.

### 4.1.2. Performance contract

The performance contract incentivizes the construction and operations for a mutually established thermal performance level during the use phase by leasing out thermal performance as a service. The smart contract directly executes payment from the escrow to the contractor and facility manager for delivering the agreed-upon performance levels. It is not the focus of this work to propose a finished performance contract, but rather to demonstrate an exemplary cross-phase incentive case that can be tested with the actual sensor data of the Tz2 building. Nevertheless, the contract is based on common thermal performance evaluation metrics.

The energy consumption of the building and the level of comfort of the building occupants are two of the most important thermal performance factors. While it is clear that high energy consumption causes increased costs, dissatisfaction of occupants regarding comfort levels also increases costs. For example, users might set up their own local heaters and coolers [87] or their work performance might decrease, leading to a rise in personnel costs [88]. Therefore, the implemented thermal performance logic measures and evaluates 1) overall energy consumption, and 2) thermal comfort levels of the building.

The logic regarding overall energy performance ($EP$, Eq. 1) compares the actual average energy consumption for a given time interval ($E_{\Delta t}$) with the expected energy consumption ($E_0$).

$$EP_{\Delta t} = \frac{E_{\Delta t}}{E_0} \tag{1}$$

Thermal comfort assessement is based on a simplified predicted mean vote model ($PMV$, Eq. 2) developed by Buratti et al. [89] based on Rohles [90], and it only relies on air temperature and relative humidity, since the original PMV model developed by Fanger [91] also takes into account air speed and mean radiant temperature, neither of which is measured in Tz2. Buratti et al. [89] provide an extensive data baseline distilled into diagrams for acceptable levels of PMV for a specific comfort scenario (determined by the parameters $a$, $b$, $c$), given the temperature ($T$) and water vapor pressure ($P_v$) derived from the relative humidity ($RH$).

$$PMV(T, P_v) = aT + bP_v - c \tag{2}$$

Based on this data, one can select a target comfort scenario and derive the required set point temperature and relative humidity.

First, the thermal comfort for room temperature ($TC_T$, Eq. 3) compares the actual average room temperature for a given time interval ($T_{\Delta t}$) to the targeted set point temperature ($T_0$).

$$TC_{T, \Delta t} = \frac{T_{\Delta t}}{T_0} \tag{3}$$

Second, the thermal comfort for relative humidity ($TC_{RH}$, Eq. 4) compares the actual average relative humidity for a given time interval ($RH_{\Delta t}$) to the targeted relative humidity ($RH_0$).

$$TC_{RH, \Delta t} = \frac{RH_{\Delta t}}{RH_0} \tag{4}$$

Finally, since $CO2$ measurements are also available in Tz2 as a good indicator of air quality, we compare the $CO2$ thermal comfort ratio ($TC_{CO2}$, Eq. 5) with the actual average $CO2$ level in a given time interval ($CO2_{\Delta t}$) with a targeted $CO2$ level ($CO2_0$) often assumed to be below 1000ppm [92].

$$TC_{CO2, \Delta t} = \frac{CO2_{\Delta t}}{CO2_0} \tag{5}$$

To summarize, the data fetched from the Tz2 sensors are indoor temperature, relative humidity, $CO2$ concentration, and energy consumption for heating and cooling. To assess the thermal performance, several factors have to be agreed on: an expected energy consumption, a thermal performance scenario determining the expected values for the set point temperature and relative humidity, and a target $CO2$ level.

*Table 1: Performance reward logic for the facility manager.*

| | Temperature | Relative Humidity | CO2 Concentration |
|---|---|---|---|
| Facility Manager | $0.9 \leq TC_{T, \Delta t} \leq 1.1$ | $0.75 \leq TC_{RH, \Delta t} \leq 1.5$ | $TC_{CO2, \Delta t} \leq 1$ |
| Facility Manager | $0.8 \leq TC_{T, \Delta t} < 0.9$ <br> $1.1 < TC_{T, \Delta t} \leq 1.2$ | $0.4 \leq TC_{RH, \Delta t} < 0.75$ <br> $1.5 < TC_{RH, \Delta t} \leq 1.8$ | $1 < TC_{CO2, \Delta t} \leq 1.1$ |
| Facility Manager | $TC_{T, \Delta t} < 0.8$ <br> $1.2 < TC_{T, \Delta t}$ | $TC_{RH, \Delta t} < 0.4$ <br> $1.8 < TC_{RH, \Delta t}$ | $1.1 < TC_{CO2, \Delta t}$ |

*Table 2: Performance reward logic for the contractor, given the facility manager's performance.*

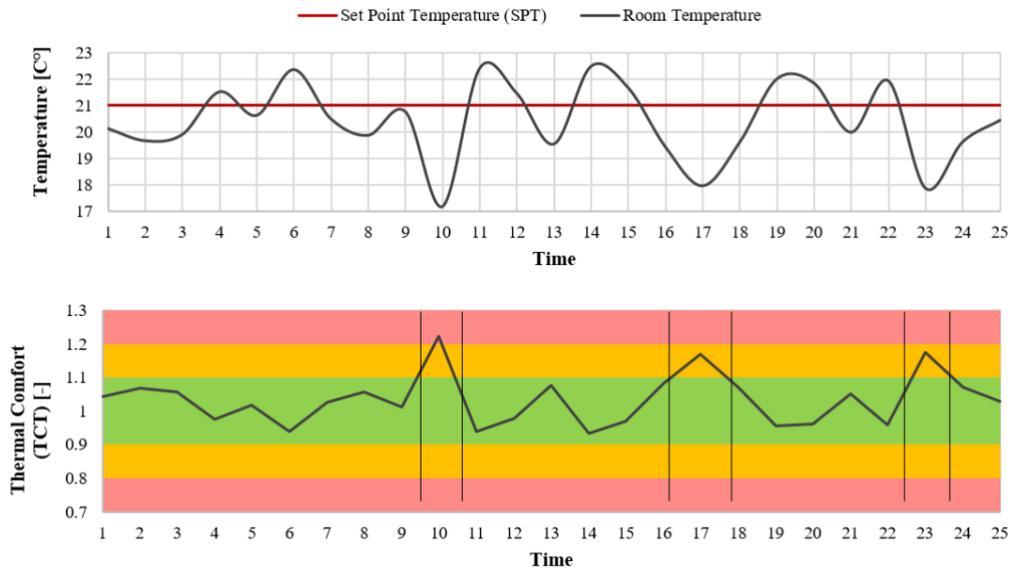

| | Energy Consumption | Facility Manger | | |
|---|---|---|---|---|
| Contractor | $EP_{\Delta t} \leq 1$ | $TC_{T,\Delta t}$ | | $TC_{T,\Delta t}$ |
| Contractor | $EP_{\Delta t} \leq 1$ | $TC_{T,\Delta t} < 0.8$ | | |
| | $1 < EP_{\Delta t} \leq 1.5$ | $TC_{T,\Delta t}$ | $TC_{T,\Delta t}$ | $TC_{T,\Delta t}$ |
| | $1.5 < EP_{\Delta t}$ | $1.2 < TC_{T,\Delta t}$ | | |
| Contractor | $1.5 < EP_{\Delta t}$ | $TC_{T,\Delta t}$ | | $TC_{T,\Delta t}$ |

*Figure 6: Exemplary visualization of performance assessment for the temperature thermal comfort ($TC_T$). For a given SPT of 21°C and the measured room temperature values (a), the $TC_T$ ratio must stay within the defined range (green), causing a reduced reward (orange) or failure (red) for the facility manager when deviating (b).*

The performance contract determines whether the contractor and facility manager succeed or fail in delivering the agreed performance levels. For the proof-of-concept, the logic assesses performance deviations in percent defined by the authors based on reasonable assumptions (Table 1, Table 2, Figure 6). The facility managers need to ensure indoor comfort, so the reward depends on reaching the expected levels for temperature, relative humidity, and CO2 concentration (Table 1). The full reward requires two out of three targets (green) to be reached. For two out of three failed targets (red), no reward is issued. In between, there is a reduced reward (orange). The contractor's reward depends on the total energy performance ratio, but in relation to the thermal comfort levels for room temperature (Table 2). This ensures both that a contractor cannot bribe the facility manager to reduce indoor comfort to meet the energy performance and that extensive heating of the building by the facility manager does not cause a failure for the contractor.

### 4.2. Technical implementation

This section describes in more detail the technical implementation of the proof of concept. An overview of the implemented components is shown in Figure 7. The code is available under an open source licence[1].

---

[1] https://github.com/mahshidmotie/PerformanceBasedSmartContracts

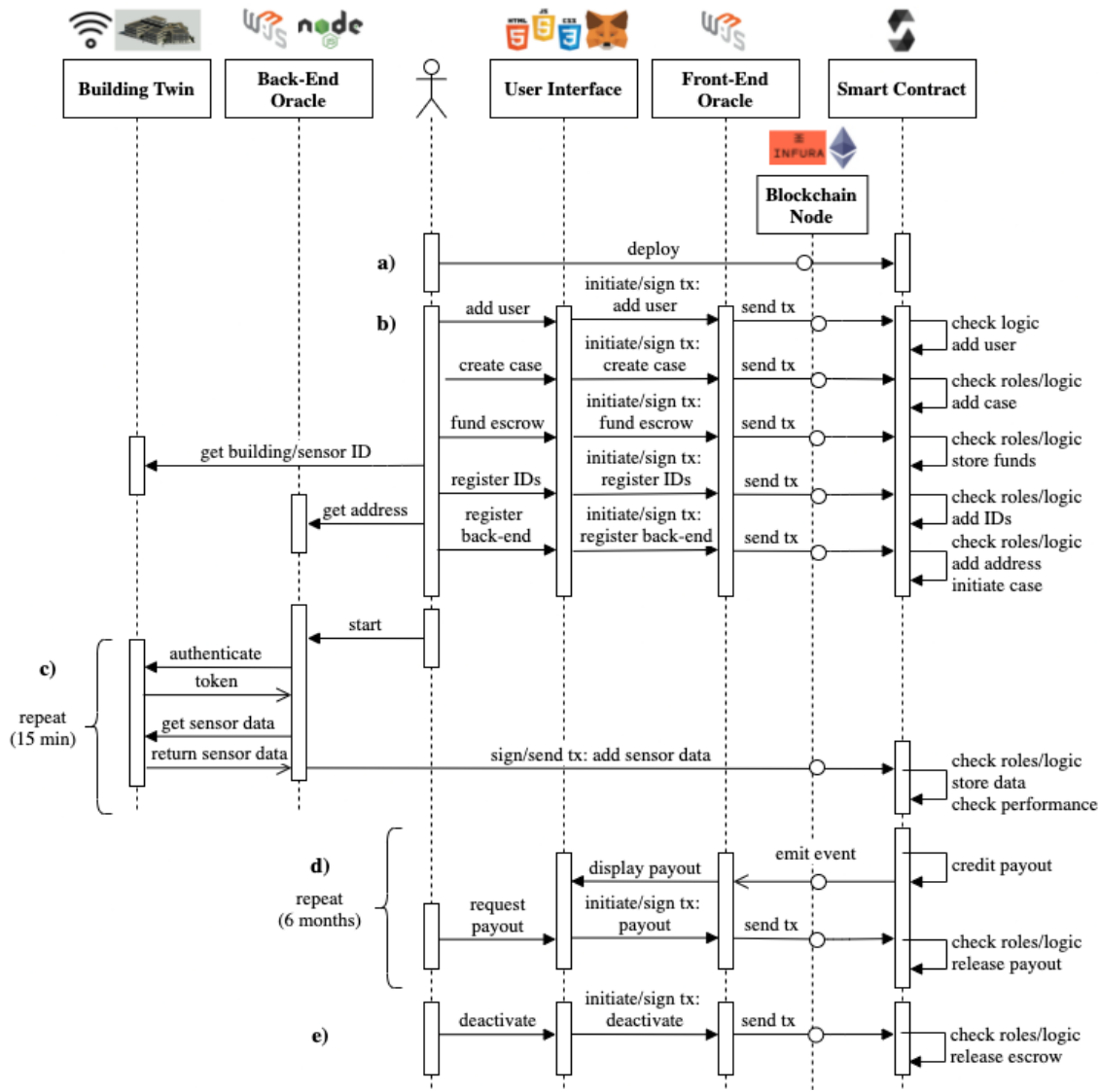

*Figure 7: Interaction of the implemented technical architecture for the proof of concept workflow.*

### 4.2.1. Thermal performance smart contract

For the scope of this proof of concept, we selected the Ethereum blockchain to develop and deploy the performance-based smart contract. At the time of conducting this research, Ethereum was the most prominent Turing-complete smart contract platform with extensive documentation available. Nevertheless, other blockchains could be chosen in the future (see 5.1.1).

The smart contract is written in Solidity, the native smart contract language of Ethereum. We developed the smart contract using the Truffle suite [92], with Ganache as a local blockchain environment (see Figure 8). The smart contract logic can be separated into two main parts: roles and access management, and the thermal performance contract logic.

Roles management for access control was implemented by inheriting the OpenZepplin [92] "roles" and "ownable" smart contract templates. In addition to the roles of the building owner, the contractor, and the facility manager, the role of the smart contract owner is important. The smart contract owner role is assigned to the person deploying the contract. This role then has the right to assign the other roles, so that they can interact with the smart contract.

The thermal performance logic in the coded smart contract functions follows the logic described in section 4.1.2. When all logic is encoded, the smart contract is deployed (see Figure 7, a)).

Afterwards, case-specific information can be added to the contract (see Figure 7, b)). First, stakeholder roles are assigned and the specific contract case is created. The stakeholders define the contract details, such as the duration, the building, the relevant sensor data, and the agreed performance baseline. Furthermore, the building owner funds and locks the escrow. This assures the other parties that funds are available and reserved for payment throughout the duration of the contract. Finally, the building and sensor IDs, as well as the address of the back-end, need to be registered in the smart contract before execution can start.

When all information is input and the contract is funded, the contract execution can begin (see Figure 7, c)). The back-end oracle is started and the defined performance data (energy and indoor-comfort data) is passed at defined time intervals from the building twin platform to the smart contract by calling the respective smart contract functions. The data is stored, the values are evaluated by the contract logic, the results are saved, and respective actions are triggered. For monetary payments, the smart contract keeps track of the amounts earned by each role.

Finally, the rewards can be redeemed through the front-end at defined intervals of 6 months (see Figure 7, d)). This interval was chosen to reduce the number of monetary transactions that need to be triggered by the stakeholders, but also other time intervals can be used. After the contract duration is complete, the building owner can release the remaining escrow amount (see Figure 7, e)).

*Figure 8: Smart contract deployment to the development network using Truffle.*

### 4.2.2. Front-end oracle

The web application is built with HTML, CSS, and JavaScript. The graphical user interface provides an input mask for the smart contract arguments to set up the smart contract or to interact with the smart contract functions. The front-end triggers transactions using the Web3.js API. To sign transactions, the user needs to use a wallet that handles the correct private keys. This ensures that only authorized roles can perform actions. In this proof of concept, we use Metamask to connect with an Ethereum node. For development purposes, we used a local blockchain instance (Ganache), but to deploy to the test network (Rinkeby), we used the Infura API to connect to remote nodes.

The contract stakeholders can use the graphical input mask in combination with Metamask to conveniently interact with the smart contract (see Figure 9). They can set up a new case, check on the contract status, and redeem their rewards.

*Figure 9: Snippet of the graphical input mask to execute the performance-based smart contract functions, using the Metamask wallet to sign transactions.*

### 4.2.3. Back-end oracle

The back-end oracle server acts as a middleware between the Siemens building twin platform and the Ethereum blockchain. It is built using Javascript and NodeJS. The connection to the Siemens building twin platform is established using its APIs. A valid access token needs to be appended to the API calls. The fetched data is then formatted and passed to the smart contract by calling the respective smart contract function using the Web3.js API and Infura API. In contrast to the front-end oracle, the same address owned by the back-end oracle always signs the transaction. Therefore, Metamask is not needed. The back-end address is registered in the smart contract, so no other address can call the function. The transactions are directly signed by the server with the private key using the web3.js wallet functionality.

Since the data is saved in the smart contract, the transaction costs increase with the number of submitted data points. Storing large amounts of data in the smart contract can be very costly and not economically viable. Therefore, the transmitted data points should be minimized without affecting the performance contract functionality. At the same time, the possibility for data manipulation (see 3.2, oracle problem) should be addressed.

Various scenarios were investigated. First, the number of sensors and therefore monitored spaces can be limited. Obviously, this would also limit data diversity since only some rooms are monitored. Moreover, a scenario with fewer sensors means a higher chance that the selected physical sensors or the sensor data in the building twin central storage could be manipulated. Second, all sensors can be fetched simultaneously, but the time intervals of fetching data can be decreased. However, specifying known and consistent time intervals poses also more attack vectors to manipulate data at exactly these points in time. The third scenario can follow a randomization strategy regarding both space and time. On average, data is fetched every 15 minutes, but with randomized variations. Moreover, at each time a random sensor is chosen out of the registered list of sensors. Finally, the number of measurements should match the number of data points needed for the evaluation logic.

After evaluation, the third scenario was selected for test implementation in this paper. In the implemented case, indoor environment measurements of sensors are selected on average five times a day, while heating energy consumption is measured only once a week.

### 4.2.4. Siemens building twin platform

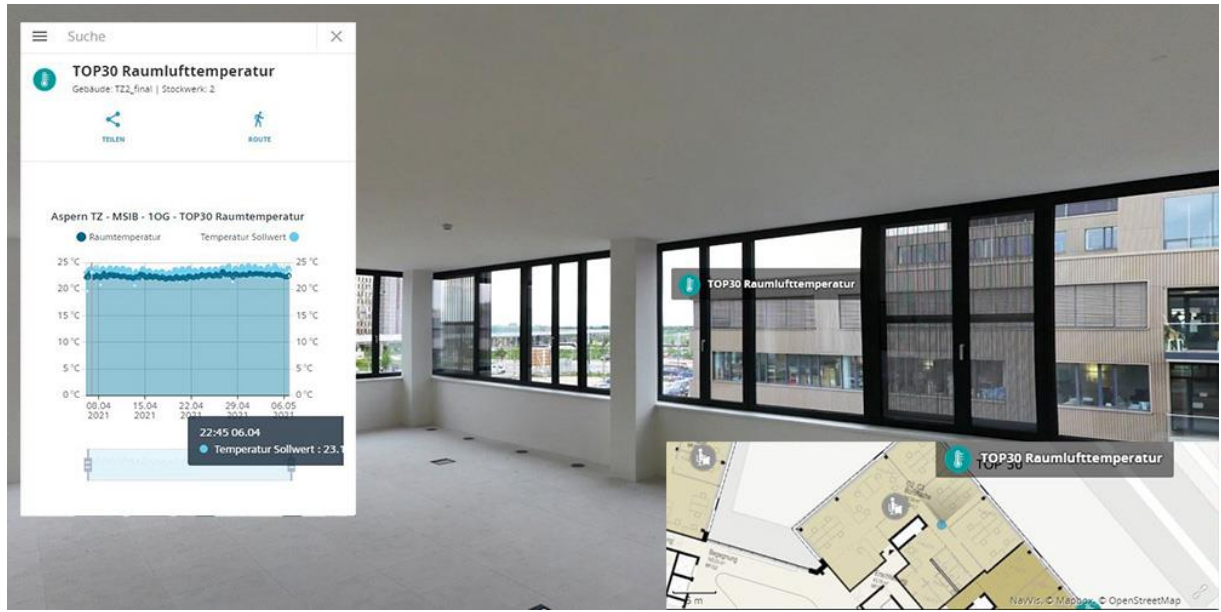

*Figure 10: The Siemens building twin platform for Tz2 (Copyright: Siemens AG).*

The Tz2 building is monitored and controlled with the Siemens building twin platform [93]. The platform is a single source for both static and dynamic data. This data is visualized in a digital 3d model by constantly updating the static building information (based on the BIM IFC files from the design and construction) with the dynamic real-time data from the connected sensors. The platform can also run performance analytics to help optimize the technical systems of the building.

In the proof of concept, we used the Siemens building twin platform of Tz2 as an external data source for the performance-based smart contracts (see Figure 10). The sensors are referenced in the static IFC files of the digital model with their BACnet addresses. This allows the sensor data to be mapped to the respective physical devices and spaces in the 3d visualization. Relevant sensors can be identified to register their BACnet addresses in the smart contract. The respective sensor data are then fetched from the digital twin database and transmitted to the smart contract.

### *4.3. Test results*

#### 4.3.1. Test Setup

To test the implemented architecture (see 4.2), the performance-based smart contract was deployed to the Rinkeby network (contract address: 0x2b8aaf9B539fA288e1dFEa8866B6b51d1cD804B3), a test network of Ethereum. An exemplary case was created with three imaginary stakeholders (building owner, contractor, and facility manager). The smart contract ran for two days starting on May 14th and ending on May 16th, 2020 on the test network.

The performance baseline for the weekly energy consumption was chosen as 45 kWh. The comfort level baselines were chosen as follows: 21° C set point temperature, 40% relative humidity, and a $CO_2$ level of 1000ppm. To cover an equivalent number of measurements as in a full winter season (6 months) within the two days, the number of thermal performance measurements was increased from 5 to 190 per day. This was needed to generate one payout event after a 6-month time interval as defined in the smart contract logic.

### 4.3.2.Transaction Data

The implemented protoype functioned as intended, validating the feasibility of the proposed architecture. By the end of the test, 1241 measurements were stored in the smart contract. All transactions were executed following the encoded transaction logic (see 4.1.2). There were several thermal performance failures observed. Further analysis revealed that the terms coded in the smart contract identified the failure correctly, so the smart contract logic worked as expected. However, it is clear that with the assumptions of the performance baseline, as well as the accelerated collection of data points, the observed performance and respective reward logic are not meaningful in terms of the actual performance of the building.

In addition, we observed the transaction costs for the test run to examine financial viability. Every transaction incurs a transaction cost paid to the miners in the blockchain network for adding the transaction to the blockchain. In public blockchains, this fee is paid in the native cryptocurrency of the network. Our prototype uses the Ethereum blockchain, so in this case the cryptocurrency is Ether (ETH). The ETH fee is calculated based on the necessary computing cost (Gas amount) for a transaction, multiplied by the Gas price determined by the current network utilization.

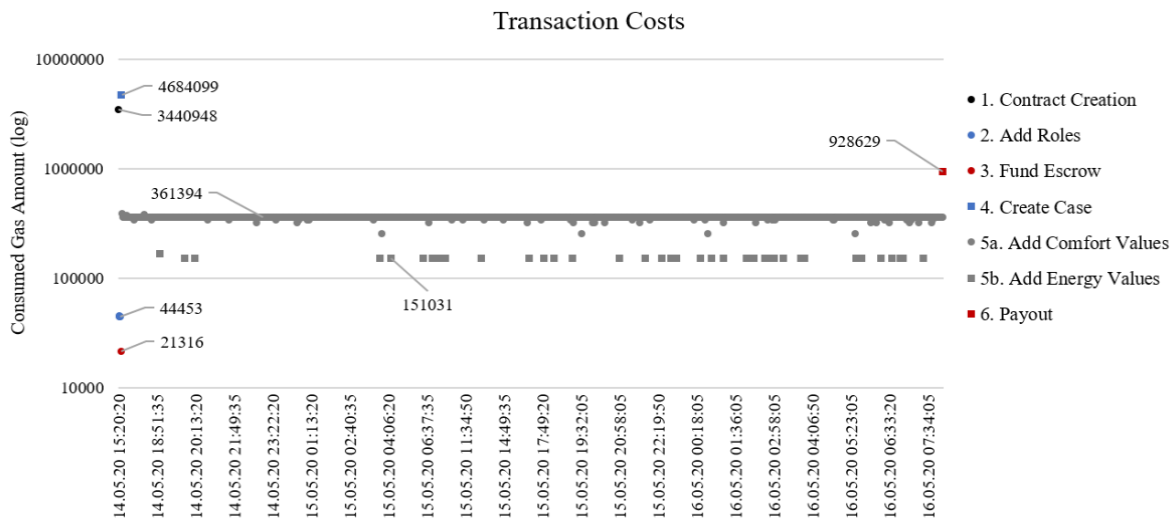

*Figure 11: Transaction costs (Gas) for the executed transactions of the test case.*

The Gas price in the a test network can be set by the developer and is therefore not meaningful. Because the protoype was deployed to the Rinkeby test network, this applied to the investigated use case. However, the Gas amount for a transaction in the test network is comparable to the Ethereum main network. Therefore, Figure 11 pictures the Gas amount for the executed transactions of the implanted contract (see also Figure 7). In the beginning, the contract was created, roles were added, a performance case was created, and the escrow was funded. For those transactions, the consumed gas amount depends on the chosen implementation of the smart contract and the number of transactions needed to pepare the contract for execution, e.g. the number of roles to be registered, or the number of transactions to fund the contract. Over the two days, the sensor data of energy and comfort values were then added according to the chosen intervals considered sufficient for the use case (see also 4.2 and 4.3.1). A final transaction calculated the rewards and released the respective payouts. The total Gas consumed by the performance contract was 460'217'196. The three most expensive transactions for the executed use case were the contract creation, case registration with all sensor IDs, and final payout calculation. Nevertheless, the cost to add the sensor data accumulated to 97% of the total transaction costs. This demonstrates the importance of reducing data stored on-chain in the smart contract for cost considerations of running performance-based smart contracts.

For an indication of expected costs in case of a real deployment to the Ethereum main net, Figure 12 shows the average historic price for the above test case (total Gas amount). The final cost in USD depends on the Gas price and the ETH price at the time of the transaction execution. Figure 12 shows the expected total cost for a six-month time period (since the tested use case would run for six months) using average Gas and ETH prices. For example, at the time of deploying the contract in the test run (May 14th, 2020), the average Gas price over the next 6 months was 89.8 Gwei, resulting in approximately 41.33 ETH total cost. With an average market price of 322.5 USD/ETH, this results in 13'327 USD. The graph reflects the impressive uptick in network use (Gas price) followed by the USD market price for ETH in late 2020, resulting in a significant increase in cost compared to the previous years.

It is important to note that the above data is highly dependent on to the presented prototype implementation using the Ethereum network and on the network state at the time of execution, as well as on the specifics of the performance based contract implemented. Costs could vary significantly using another blockchain or a different use case scenario.

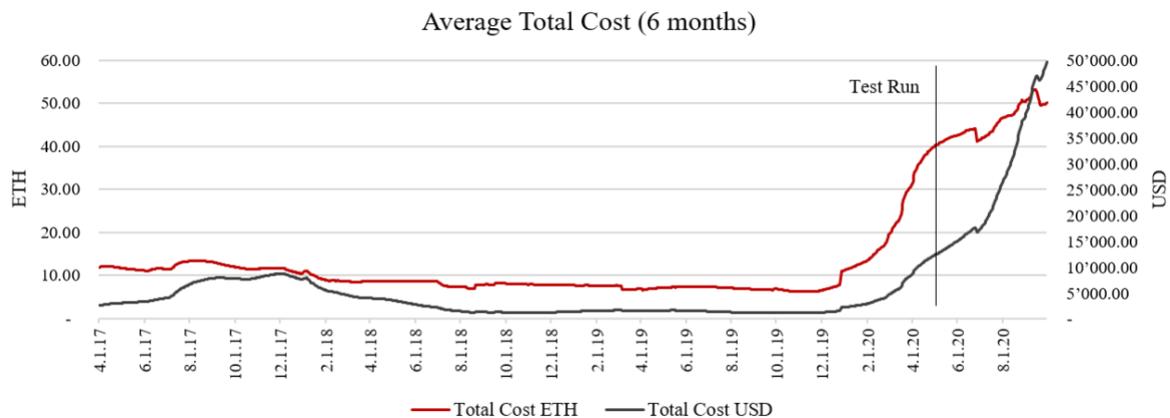

*Figure 12: Total costs if deployed to the main net based on a six-month average Gas and ETH price after the time of deployment (Data source: [94,95]).*

### 4.3.3. Stakeholder Feedback

A short survey collected stakeholder feedback on the concept of performance-based building using digital building twins and blockchain smart contracts after showcasing the prototype. Figure 13 shows the benefits and challenges mentioned by nine stakeholders, sorted according to the number of mentions. The small sample size has no statistical significance, but we found it nevertheless helpful to see the perceived benefits/challenges and to cross-check them with our own assessment.

Overall, the stakeholders demonstrated interest in using a (more mature) solution based on blockchain and digital building twins and had general confidence that it could be successful in introducing new incentives towards better performance and more efficient buildings (Figure 13, b). If stakeholders referred to the technical solution, they found the automated and verifiable approach especially appealing.

On the other hand, many challenges and concerns were mentioned (Figure 13, c). The concern that was most often mentioned was that performance-based building will change processes so that they are no longer compatible with existing business relations. This is somewhat surprising, since the inherent idea of performance-based building is in fact to change business processes (see 2.1) and provide the respective incentives to make these changes be perceived as a benefit. Moreover, among the other top mentioned challenges were the definition of fair performance evaluation criteria, accurate energy performance simulations to determine the expected performance baseline, and legal limitations. Interestingly, none of these is related to the technical system but rather to general barriers to performance-based building. The most often-mentioned technical challenge was the development and maintenance of digital building twins, followed by the technical security and maturity of both digital building twins and blockchain, and then followed by the shift in trust to the technical system.

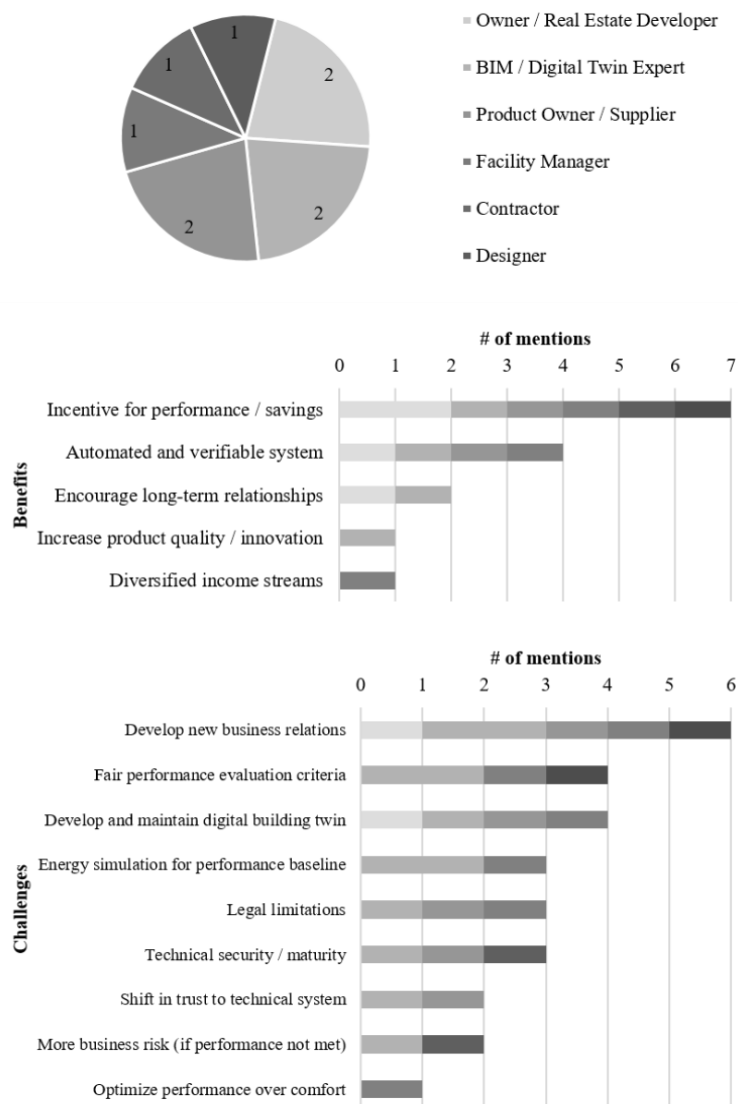

*Figure 13: Survey results. Participating stakeholders (a), mentioned benefits (b), and mentioned challenges (c).*

**5. Discussion and outlook**

### *5.1. Proof of concept*

In this paper, we present what is to our knowledge the first full-stack prototype for a performance-based smart contract in the built environment. To do this, we integrate the Ethereum blockchain with digital building models and sensors via the Siemens building twin platform. The successful proof-of-concept shows the feasibility of both the concept and implemented technical architecture. Nevertheless, we found that as emerging technologies, both digital building twins and blockchain need to mature for scalable and secure real-world implementation. The following discussion structured according to the different technical components (see Figure 2) identifies the limitations we observed as well as relevant considerations for future research.

5.1.1. Blockchain

The proposed use case falls at the intersection of three proposed use cases for blockchain in construction [63]. It is an example of "coins/tokens as payment or incentive scheme across the whole life-cycle" for the performance of a built asset, combined with "transaction automation with smart contracts" for automatic evaluation of performance and contract terms and with "immutable and transparent records of transactions" to the facilitate trust of participating stakeholders in its actual execution.

Even though this proof of concept used the public permissionless blockchain Ethereum, the question of which DLT option best fits the proposed use case can be further assessed and debated. According to the proposed classification in Hunhevicz and Hall [63], the above categories could use different DLT options, depending on whether the participating stakeholders are known and whether public verifiability is desired. Currently, all stakeholders are generally known and companies are mostly skeptical towards public verifiability. Therefore, in the short term, private permissioned blockchains can be attractive for more network control and privacy. In the future, new servitized business cases might emerge that promote long-term incentive mechanisms that need to be set up without knowing all potential stakeholders at the time of setting up the service contract. This would shift preferences towards public permissionless DLT systems. Permissionless DLTs are more decentralized and robust networks are likely to exist also in decades to come, whereas the permissioned networks might rise and fall with the central entities controlling the network.

Furthermore, the use of cryptocurrencies is a strong argument for public permissionless blockchains to assure long-term trust [63]. While this could be bypassed in the short term by connecting to legacy payment systems, using cryptocurrencies reduces effort when relying on smart contracts for the performance contract. However, as demonstrated in this study, reliance on cryptocurrencies such as Ethereum can suffer from both ETH price volatility and network congestion rates resulting in high Gas prices (see Figure 12). While ETH price volatility could be addressed through the use of stable coins or DEFI future contracts, high network use driving Gas costs is a concern for long-term contracts as proposed in this paper. This might be resolved with further advances in technology or alternative DLTs, but it nevertheless shows the importance of writing efficient code that reduces on-chain computation and storing as little data as possible on-chain. Both of these could have been optimized in our implementation. Further assessment is needed to determine if this optimization would suffice to obtain the price levels required for greater industry adoption.

Overall, public permissionless blockchains seem like a good future fit for the use case, even though the shortcomings of current public permissionless DLTs (e.g. throughput, privacy, transaction costs) need to be addressed for large-scale implementation. In the short term, it could make sense to start with more scalable and cheaper private permissioned DLTs to test performance-based smart contracts in a real business setting and move with more technical advancements towards public permissionless DLTs. However, further research with different DLTs should be conducted to provide more nuanced insights.

### 5.1.2. Performance-based smart contract

The implemented cross-phase thermal performance contract demonstrated an exemplary smart contract implementation in Solidity. The smart contract functionality worked as expected in the two-day test run. Nevertheless, the proof of concept revealed many challenges and limitations that should be addressed in future research.

The implemented thermal performance contract logic is very preliminary. The workflow and participants involved were simplified for demonstration purposes. Moreover, the thermal performance evaluation needs to be refined. Also, the payouts were chosen randomly – no appropriate rewards for the given business case were calculated.

To move the field of performance-based smart contracts further, more research needs to first assess the suitable logic and incentives for cross-phase performance-based contract terms. We encountered many questions when setting up the contract logic. Does only the owner need to pay an escrow or do all stakeholders need to lock funds to demonstrate skin in the game? Should participants only be rewarded or also punished if the target is not met? How is performance measured fairly and how can cheating be avoided? How can external effects (e.g. weather) be excluded? What is a fair price for a service provided? Overall, valid business cases need to be established as servitization use cases, most importantly the fair performance baselines and rewards. This was also mentioned as an important challenge in the stakeholder survey (see Figure 13). Finally, from the exemplary prototype in this research, it is not yet clear whether the performance-based smart contracts with the presented technology stack can be applied to all aspects of building performance.

The smart contract can be coded only when the contract logic is defined, and for this simple proof-of-concept, the Solidity language was sufficient to encode the terms. However, it became apparent that Solidity has its limitations when trying to implement advanced mathematical calculations. Furthermore, experts should be consulted to make sure there are no security issues that could lead to the loss of funds. Once the smart contract is deployed, it is very hard or even impossible to patch ex-post when no governance mechanism for such adjustments was implemented beforehand. Therefore, ensuring the flexibility of smart contracts in handling unexpected cases will likely be a major challenge. It is important not to erode the advantages of smart contracts by implementing admin functionalities that again introduce third-party risk (e.g. to halt the contract). Gürcan et al. [85] proposed establishing agreed-upon processes on how to encode smart contracts. Ultimately, a smart contract could be assembled based on modular pieces that automatically comply with legal terms. But this was not further assessed in this research. Future research needs to investigate the legal and regulatory situation and challenges within different jurisdictions when trying to implement the proposed performace based smart contracts. This was also mentioned repeatedly by the interviewed stakeholders as a challenge (see Figure 13).

Furthermore, the storage of data poses major challenges. In the proof-of-concept, fetched sensor data was stored within the smart contract. This causes increased transaction costs and potential issues with the privacy of data in public blockchains, and it strains the network through blockchain bloat.

We identified different approaches to on-chain and off-chain data storage. First, as done in our proof of concept, the number of stored measurements could be decreased through the randomization approach. Nevertheless, data stored on-chain still aggregates over time to sizable amounts. As a possible alternative, performance metrics could be calculated off-chain from externally stored sensor data and only aggregated information stored in the smart contract. This might provide an even better balance between trusted execution of critical functions in the smart contract (final reward decisions) and storing large amounts of data off-chain. Lastly, no performance data could be stored and calculated on-chain. Data sent on-chain would only include whether performance was met (true/false) from the digital building twin to initiate payments. A tradeoff remains between more trust but more expensive on-chain data storage, or off-chain data storage but less trust (see also 5.1.3). Future research should assess further possibilities for harmonization and preprocessing of data before saving in smart contracts together with the implications for overall trust in the solution.

### 5.1.3. Digital building twin as external data source

In the proof of concept, the building twin is used as an intermediary platform that connects to the sensors and stores sensor data. The advantage is the ease of data access, the possibility to select sensor data, and the potential for harmonization of data upfront. Overall, the digital building twin reduces the amount of data that must be stored in the smart contract. The disadvantage is that this introduces a dependency on a centralized third-party service with a potential single point of failure (e.g. the building twin provider cease operations). The randomization approach implemented here to fetch data addresses some of the potential attacks that could manipulate data, but it does not eliminate the dependency on the building twin platform Moreover, cross-phase performance contracts also require that a building twin is available and maintained across all life-cycle phases of the built asset. This is still a challenge (see stakeholder feedback in Figure 13) and rarely achieved nowadays, which considerably limits the number of built assets to which the proposed architecture can currently be applied. To reduce dependency on digital building twins, sensor data could be fetched directly from the sensors into the smart contract, so the only prerequisite for the built asset is that it is equipped with the relevant sensors. However, this solution would again complicate efforts to clean and process data and cause problems with data storage on-chain.

Furthermore, connecting the blockchain with the digital building twin and directly to the sensors relies on a back-end oracle. Since this single point of failure is critical to the functioning of the system, the use of a centralized digital building twin platform is, in the view of the authors, acceptable in the near term. Future research could investigate how the single points of failure described here could be addressed, e.g. through implementing decentralized server meshes.

Overall, implementing a secure back-end is challenging and requires further research. The convenience of using digital building twin platforms comes with a tradeoff in security and redundancy that could affect trust in the whole system but might be necessary to reduce on-chain data storage. Besides ensuring a secure technical infrastructure, future studies need to look into additional security layers to combat the potential impact of human factors (e.g. fraud) when interacting with the BIM models, digital building twin platform, or physical sensors [73], as well as with the blockchain itself [96].

### 5.1.4. Front-end

The front-end application ensures that stakeholders can set up and interact with the smart contract. Therefore, it is a critical piece of infrastructure to make the solution as simple to use as possible to overcome socio-technical barriers [61]. More research should investigate easy-to-use front-end applications that provide functionality for setting up and interacting with performance-based smart contracts. As for the back-end side (see 5.1.3), security issues caused by human factors should also be examined for the front-end oracle.

### 5.2. Crypto economic life-cycle incentives for servitization

It was found that presenting performance-based building as a compelling business case rather than a technical issue can be one of the main enablers to performance-based building [14]. This proof-of-concept has provided insight into the potential of using performance-based smart contracts for a future servitized built environment. The use case scenario demonstrates the potential of crypto-economic incentives to align performance targets for new profitable business cases without relying on any trusted third party, and as a side effect benefit the environment by saving energy and reducing $CO_2$ emissions.

Benefits could increase with more advanced servitization business cases. Smart contracts enable scalable collaboration between many parties with low bureaucratic overhead by continuously saving transactions transparently in the blockchain coupled to automated reward logic. Also, the possibility of coding incentive systems through tokens has not been assessed in this paper. Next to crypto currencies (money) for payments, other reward tokens could be issued for reputation or non-monetary performance metrics, e.g. environmental impacts [97,98]. Such new crypto-economic life cycle incentives could motivate further business cases. These incentives could move the built environment towards servitization between anonymous stakeholders, enabled by the trust provided by performance based smart contracts. Producers and owners might provide their built assets with publicly available service contracts on the blockchain, while other service providers and users can evaluate available offers and directly sign these contracts on the blockchain, getting paid for their performance or paying anonymously and peer-to-peer for the service used.

## 6. Conclusion

The combination of blockchain-based smart contracts with digital building twins is promising to 1) digitize performance contracts in a trusted way and scale performance-based use cases in the built environment, and 2) enable new business models through crypto-economic incentives linked to the life-cycle performance, which might motivate more stakeholders to explore a built environment as a service.

Feasibility of the above was demonstrated with the first full-stack proof-of-concept of an exemplary thermal performance-based smart contract, using the Ethereum blockchain and the Siemens building twin platform connected to the sensors of a real-world building. The early technical infrastructure is available. Nevertheless, many limitations apply. We see the main contributions of this paper as pointing out the challenges that require further research.

Despite the positive feedback of stakeholders regarding the potential of the solution, a major challenge will be to define a fair logic for performance-based contracts and performance baselines. This is also what the authors observed when setting up the thermal performance smart contract: coding the contract logic was relatively straightforward compared to the challenge of defining performance logic and respective payouts. Smart contracts are an emerging tool to realize more scalable and attractive performance contracts, but more research needs to first investigate the underlying performance logic and associated business models.

The early state of blockchain leads to many technical challenges that need to be addressed for scalable and secure implementation of performance-based smart contracts. Currently, the usability of blockchain infrastructure is not at the required level to protect stakeholders from errors when setting up and interacting with the contracts. While in the early years of Ethereum the observed costs were reasonable for the tested performance-based smart contract, the recent price and network use increase have led to unreasonable price levels that need to be addressed for large-scale implementation. A related challenge is how to reduce on-chain data storage without compromising the trust provided by the smart contract. Furthermore, while digital building twins simplify the connection of smart contracts to the real-time performance data of the building, the precise interaction needs more research. A secure interplay between centralized infrastructure and the trusted and decentralized blockchain environment is not straightforward. Finally, the proposed solutions rely heavily on well-developed and maintained digital building twins. As of now, this is not often pursued or achieved in the industry.

The paper demonstrates the potential of the interplay between blockchain and digital building twins for performance-based smart contracts to leverage crypto-economic incentives in moving towards a trusted peer-to-peer economy in a built environment as a service. This combination can align incentives for better performance with a smaller environmental footprint, while still allowing for profitable business cases.

## 7. Author contributions

JH and MM contributed equally to this work. JH conducted the writing and data analysis for this article. MM implemented the code and wrote the master thesis that acted as a base for this paper under the supervision of JH and DH. DH contributed to the background of performance-based contracts, materials and publications for citation, feedback, and direction for content. All authors helped to finalize the article and approved the submitted version.

## 8. Funding

This research did not receive any specific grant from funding agencies in the public, commercial, or not-for-profit sectors.

## 9. Acknowledgment

The authors thank Dr. Berit Wessler, Markus Winterholer, and Oliver Zechlin of Siemens Smart Infrastructure for their support and access to the Siemens building twin platform, and the Aspern smart city research project for access to the sensor data. Finally, the authors also thank the nine industry experts that provided their insights and feedback to the solution.